\documentstyle[epsfig,epsf]{article}
\newcommand{\ndt}{\noindent}
\newcommand{\und}{\underline}

\begin{document}

\begin{center}
{\Large {\bf The treatment and rejection of two ``strange" candidate events in the SLIM experiment}}

\vskip .4 cm

S. Balestra$^{1,2}$, 
S. Cecchini$^{1,3}$, 
M. Cozzi$^{1,2}$,
L. Degli Esposti$^2$,
D. Di Ferdinando$^2$, 
M. Errico$^{1,2}$,
F. Fabbri$^2$,  
G. Giacomelli$^{1,2}$, 
M. Giorgini$^{1,2}$, 
G. Grandi$^2$, 
A. Kumar$^{1,4}$,
J. McDonald$^5$,
G. Mandrioli$^2$, 
S. Manzoor$^{1,6}$, 
A. Margiotta$^{1,2}$,
E. Medinaceli$^{1,7}$, 
L. Patrizii$^2$, 
J. Pinfold$^5$,
V. Popa$^{2,8}$, 
I.E. Qureshi$^6$,
O. Saavedra$^{9,10}$,
G. Sirri$^2$, 
M. Spurio$^{1,2}$,
V. Togo$^2$,
C. Valieri$^2$,
A. Velarde$^7$ and
A. Zanini$^{10}$ \par~\par

{\small\it 
$^1$Dip. Fisica dell'Universit\'a di Bologna, 40127 Bologna, Italy\\
$^2$INFN Sez. Bologna, 40127 Bologna, Italy\\
$^3$INAF/IASF Sez. Bologna, 40129 Bologna, Italy \\
$^4$Physics Dept, Sant Longowal Institute of Eng. and Tech., Longowal 
148 106, India \\
$^5$Centre for Subatomic Research, Univ. of Alberta, Edmonton, Alberta 
T6G 2N4, Canada\\
$^6$PD, PINSTECH, P.O. Nilore, and COMSATS-CIIT, No. 
30, H-8/1, Islamabad, Pakistan\\ 
$^7$Laboratorio de F\'{i}sica C\'osmica de Chacaltaya, UMSA, La Paz, Bolivia\\
$^8$Institute of Space Sciences, 077125 Bucharest-M\u{a}gurele, Romania\\
$^9$Dip. Fisica Sperimentale e Generale, Universit\'a di Torino, 10125 
Torino, Italy\\
$^{10}$INFN Sez. Torino, 10125 Torino, Italy\\
} 
\end{center}

\vskip .7 cm
{\bf Abstract.}
{\normalsize 
During the analysis of the CR39 Nuclear Track Detectors (NTDs) of the
SLIM experiment exposed at the high altitude lab of Chacaltaya
(Bolivia) we observed a  
sequence of puzzling etch-pits. We made a detailed investigation of
all the CR39 and Makrofol  
detectors in the same stack and in all the stacks around the candidate event. 
We found a second puzzling sequence of etch-pits (plus some single etch-pits). 
 The analysis of this configuration was important because we were
 searching for rare particles (Magnetic Monopoles, Nuclearites,
 Q-balls) in the cosmic radiation. Thus we analyzed  
in detail the evolution with increasing 
etching time of the etch-pits. We concluded that the two sequences of
the etch-pits (and some other background etch-pits) may have originated from a rare manufacture malfunctioning which involved 1 m$^2$ of produced CR39.}

\large
\section{Introduction}
In the past, a number of magnetic monopole [1-3] and 
exotic physics [4-7] candidate events 
were observed and some results were published \cite{NYtime}. 
But these results were not 
confirmed and most of them are now neglected. \par
\begin{figure}[!ht]
 
\centerline{\epsfysize=11cm\epsfbox{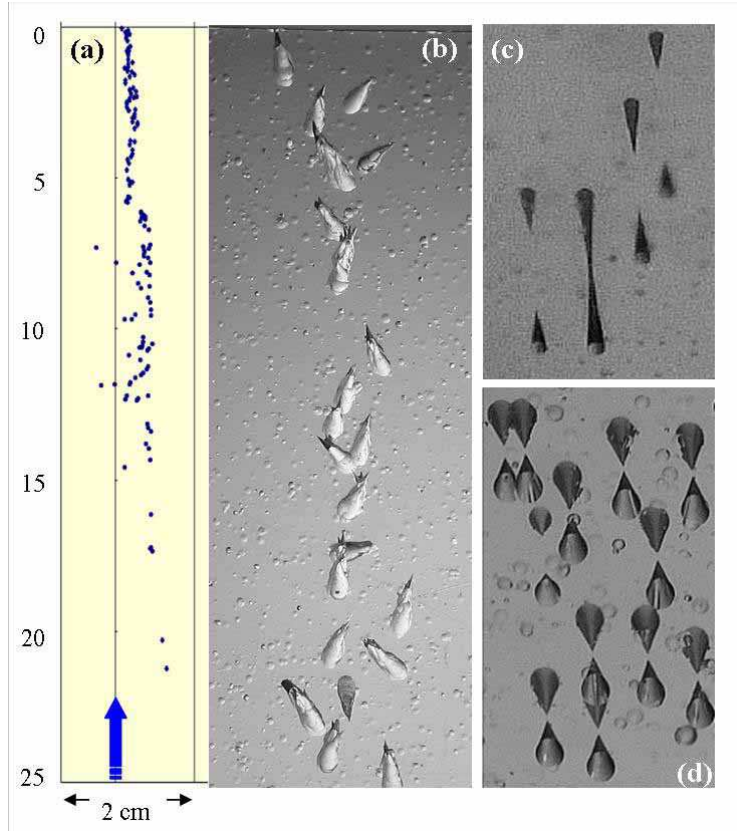}}
 \vspace{-0.5cm}
  \begin{quote}
  \caption{\small (a) Global view of the ``event'' tracks in
  the L1 layer (top surface) of wagon 7408 exposed at Chacaltaya from 20-2-01 
to 28-11-05 (strongly etched on 9-5-06), (b) Microphotographs of the
  22 tracks at the top of (a). (c) Tracks of 158 A GeV
  Pb$^{+82}$ ions and their fragments from a CERN-SPS exposure (soft
  etching), and (d) of 400 A MeV Fe$^{26+}$ ions and their fragments
  from the HIMAC, Japan (strong etching).}  
\label{fig:tracce} 
\end{quote} 
\end{figure}

In 2006 we observed an unexpected signal, analyzing the upper face of the top CR39 layer of stack 7408 of the SLIM experiment \cite{slim}. We observed a sequence of etch-pits (``event'') on the surface along a 20 cm line; each of them looked complicated (and possibly with many prongs), see Figs. \ref{fig:tracce}a,b. For comparison Figs. \ref{fig:tracce}c,d show normal tracks from 158 A GeV Pb$^{82+}$ and 400 A MeV Fe$^{26+}$ exposures at the CERN-SPS and at the HIMAC in Japan, respectively.

 A program of cross checks was started, analyzing all the CR39 and Makrofol sheets in stack 7408, and then in 48 stacks surrounding it, as shown in Fig. \ref{fig:tabella}.

We found a second sequence of etch-pits (``event'') in the bottom layer (top face) of module 7410 and some isolated etch-pits in other modules.

Importantly, the ``tracks'' evolution of this second candidate with continuing etching cast further doubts about it. We also found a few single etch-pits which, after a ``soft'' and ``strong'' etching, look like those shown in Fig. \ref{fig:tracce}b. \par
These two abnormal event candidates were of a kind never seen before. It has to be recalled that the MACRO underground experiment 
\cite{Ambrosio,riv.NC} used NTDs similar to those of SLIM.
We analyzed the MACRO CR39 detectors without finding any similar strange event candidate in a total surface area of $\sim$1200 m$^2$; the MACRO NTDs were subject to much lower ambient background levels. \par

 In this paper we report on the analysis of these ``strange" event candidates and the conclusions reached. We hope that it is of interest to researchers in the field and in the interpretation of similar phenomena observed in previous experiments.

\section{Experimental procedure}
The SLIM detector consisted of an array of NTD modules with total sensitive area $\sim$427 m$^2$. Each module was made of 3 layers of CR39 (L1, L3, L6), 3 layers of Makrofol (L2, L4, L5), 2 layers of Lexan (L0, L7) and 1 mm layer of aluminum absorber. The area of each NTD layer was 24cm $\times$ 24cm. The SLIM experiment was exposed for $\sim$4.22 years at the Chacaltaya laboratory in Bolivia \cite{slim}, 5320 m above sea level.

{\small
\begin{center}
\begin{tabular}
{|c|c|c|c|c|c|c|}
\hline
 & & & & & & \\
~7247~ & ~7248~ & ~7249~ & ~7250~ & ~7251~ & ~7252~ & ~7253~ \\
 & & & & & & \\
\hline
 & & & & & & \\
7313 & 7314 & 7315 & 7316 & 7317 & 7318 & 7316 \\
 & & & & & & \\
\hline
 & & & & & & \\
7339 & 7340 & 7341 & 7342 & 7343 & 7344 & 7345 \\
 & & & & & & \\
\hline
 & & & & & & \\
 7405 & 7406 & 7407 & {\und {\bf7408}} & 7409 & {\und {\bf7410}} & 7411 \\
 & & & & & & \\
\hline
 & & & & & & \\
7431 & 7432 & 7433 & 7434 & 7435 & 7436 & 7437 \\
 & & & & & & \\
\hline
 & & & & & & \\
7457 & 7458 & 7459 & 7460 & 7501 & 7502 & 7503 \\
 & & & & & & \\
\hline
 & & & & & & \\
7523 & 7524 & 7525 & 7526 & 7527 & 7528 & 7529 \\
 & & & & & & \\
\hline
\end{tabular}
\label{fig:tabella}
\end{center}
\vspace{0.5cm}
\begin{center}
\begin{quote}
{\small \ndt Figure 2: Layout of the SLIM modules near module 
7408 at Chacaltaya; 
each module is of 24cm$\times$24cm. Note that in the L1 layer (top face)
of module 7408 we found the first ``event'', while the second one was 
found in the L6 layer (top face) of module 7410.}
\end{quote}
\end{center}
\vspace{0.5cm}
}

After exposure the full 427 m$^2$ of NTDs was analyzed. The normal analysis procedure was as follows.

 First, the top CR39 foil of each module was etched in a 8N KOH + 1.5\% ethyl alcohol solution at 75 $^{\circ}$C for 30 hours.
The addition of alcohol improves the surface quality of the detector; but, it raises the detector threshold to $Z/ \beta \simeq 14$ \cite{slim, balestra}, corresponding to REL (Restricted Energy Loss) $\sim$200 MeV cm$^{2}$ g$^{-1}$. The scanning was done with a stereo binocular microscope with a global magnification $G=16 \times$. 

Second, if a track candidate was found we etched in a 6N NaOH + 1\% ethyl alcohol solution at 70 $^{\circ}$C 
for 40 hours the bottom L6 CR39 layer and eventually the intermediate foils. These foils were analyzed with a microscope with higher global magnification (500$\times$) \cite{slim}. A good candidate event would exhibit aligned etch-pits in all three CR39 foils. The Makrofol was planned to be used for confirmation.

 We observed the first strange candidate event on the top surface of the first CR39 sheet (L1) of the SLIM module 7408 after strong etching.  

The global view of the $\sim$140 etch-pits (``tracks'') comprising the 
first strange event candidate shows that the ``tracks'' lie in a narrow 
band roughly 2 cm wide and 20 cm long. The first 22 
``tracks'' (top part of Fig. \ref{fig:tracce}a) are shown in 
Fig. \ref{fig:tracce}b. 
	
The measurements of the etch-pit projection length ($Lp$) 
and depth ($Dp$) parameters, defined in Fig. \ref{fig:sketch}, were 
made using a microscope in transmission mode at 
$\times20_{ob}$ $\times10_{oc}$ magnification 
\cite{balestra, manzoor}. The etch-pits length ($L$) and zenith 
angle ($\theta$) were determined using the following relations

\begin{equation}
L = \sqrt{(Lp)^2 + (Dp)^2}~~~~ ,~~~~ \theta = [tan^{-1} (Lp/Dp)] 
\end{equation}
\\
where the lengths are in microns (with an accuracy of $\pm 1~\mu$m) and the angles in degrees. 

 The refractive index $n$ of the etched sheets was determined by the ratio of the thickness measured with a microscope, by focusing on the top and bottom surfaces of the sheet, and the thickness measured by a micrometer gauge (TESA). 
 We found $n=1.6$. 

\setcounter{figure}{2}
\begin{figure}[!ht]
\begin{center}
\mbox{\epsfig{figure=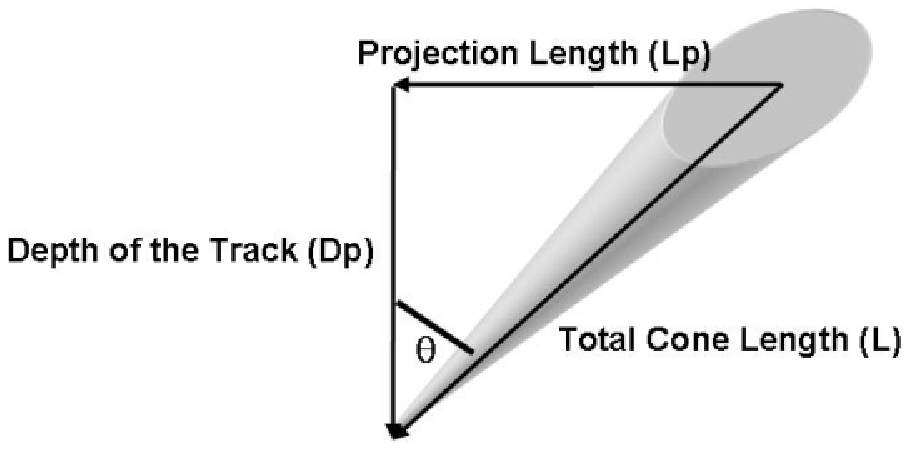,height=4.5cm}}
\caption {Sketch of the measured track parameters.}
\label{fig:sketch}
\end{center}
\end{figure}

{\it Strong etching} \cite{balestra}. The thickness of the CR39 sheets was measured in 9 different positions in the detector foil before and after etching. The average original thickness of the L1 sheet of stack 7408 was 1.4 mm; 220 microns were removed from each side of the sheet at the rate of 7.2 $\mu$m/hr for an etching time of 30 hours. The final average thickness after etching was $\sim$1 mm. 
The etching was performed in two large 65 l, temperature controlled, tanks in which 25 detectors were etched at one time. 

{\it Soft etching} \cite{balestra}. It was performed at a 
lower temperature, 70 $^{\circ}$C, with 6N NaOH + 1\% ethyl alcohol for a longer time, $\sim$40 hours. 

Measurements of strange event candidate ``tracks'' were made manually with a microscope and also with a fast 
scanning system comprised  of: (i) a low magnification  optical system 
mounted on a motor driven vertical stage for focusing; (ii) a motor 
driven scanning stage for horizontal (X-Y) motion; (iii) a digital camera 
for image grabbing interfaced to a PC for image processing; and, (iv) a light source located on the head of the optical system for illuminating the sample. Two types of images were taken: 
(i) ``wide field" images of the front surface of a NTD sheet; and, (ii) 
images from the ``inside'' the plate. A visual library formed from these 
images was used for making observations and also for making histograms of 
various important quantities, such as the track direction and length. The average track length is $\bar{L} \simeq$651 $\mu$m, and the average zenith angle is $\bar{\theta} \simeq$38 $^\circ$. 

\begin{figure}[!ht]
\begin{center}
\mbox{\epsfig{figure=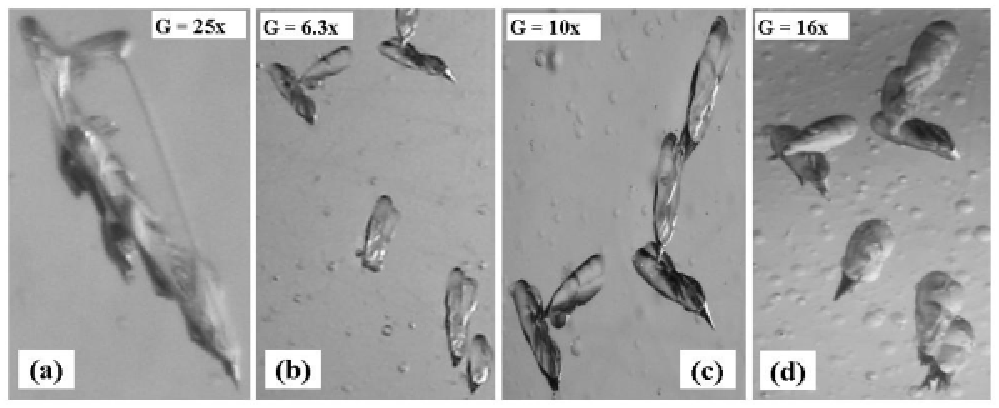,height=5cm}}
  \begin{quote}
  \caption{\small Example of ``tracks'' in the L6 layer (top face) of module 7410: (a) after 30 hours of soft etching, (b) after 5 more hours of strong etching, (c) after 4 hours of more strong etching and (d) after 10 hours of more strong etching. In each photo the global magnification $G$ used is specified.}
  \label{fig:newevent} 
\end{quote} 
\end{center}
\end{figure}

\begin{figure}[!ht]
\begin{center}
\mbox{\epsfig{figure=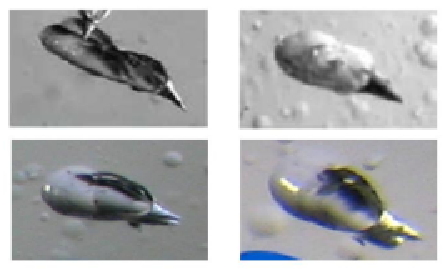,height=5cm}}
  \begin{quote}
  \caption{\small Micro photographs of selected tracks of the strange candidate events at a global magnification $G=16 \times$ in module 7408 (bottom figures) and in module 7410 (top figures).}
  \label{fig:micro} 
\end{quote} 
\end{center}
\end{figure}

\begin{figure}[t]
\begin{center}
\mbox{\epsfig{figure=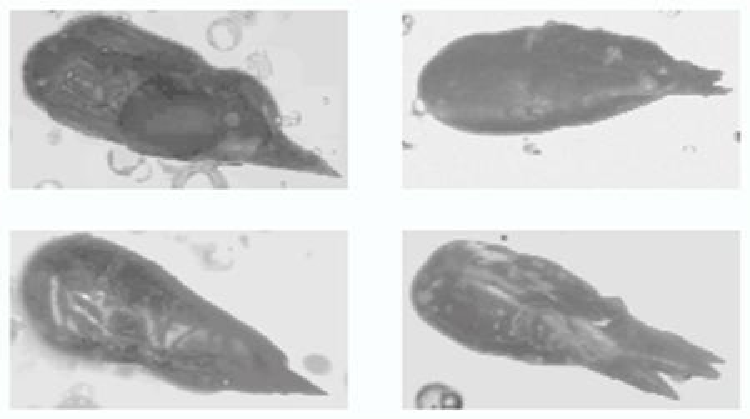,height=5cm}}
\begin{quote} 
\caption{\small Micro photographs of selected tracks of the strange events made with the fast scanning system in module 7408 (bottom figures) and in module 7410 (top figures).}
\label {fig:pic}
\end{quote}
\end{center}
 \end{figure}

 Examples of ``tracks'' analyzed with the stereo microscope are shown in Figs. \ref{fig:tracce}, \ref{fig:newevent}, \ref{fig:micro} and with the 
fast scanning system in Fig. \ref{fig:pic}. The photographs made with the fast scanning system are of lower quality, but confirm the results obtained with the stereo microscope, and yielded an easily accessible visual library.

 CR39 and Makrofol detectors, from the same batches used in SLIM, were exposed to: 1 A GeV Fe ions from the Alternating Gradient Synchrotron (AGS) at the Brookhaven National 
Laboratory (BNL); 158 A GeV In$^{49+}$ and Pb$^{82+}$ ions at the CERN Super Proton 
Synchrotron (SPS); and to 400 A MeV Fe$^{26+}$ ions at 
the Heavy Ion Medical Accelerator (HIMAC) in Japan \cite{balestra, manzoor}.
 The thickness of each sheet was measured in 25 positions in the detector foil before and after etching. The etched detectors were measured with an automatic image analyzer system \cite{noll}. The tracking was performed using the front faces of two or more sheets for  calibrations and for fragmentation studies 
\cite{dekhissi}.

\section{The First Strange Event Candidate} 

The first strange event candidate in the L1 CR39 layer (top face) in module 7408, is shown in Fig. \ref{fig:tracce}.
Two tracks from this event are also shown in Fig. \ref{fig:micro} bottom. The tracks have strange shapes and seem to be made up of several prongs with some tracks ending their range normally while others have sharp cones at their endings. There were no tracks in the L3 and L6 CR39 sheets nor in the Makrofol L2, L4, L5 sheets.

\section{The Second Strange Event Candidate} 

The second strange event candidate, in the L6 CR39 layer 
(top face) in module 7410, is shown in Figs. \ref{fig:newevent} and 
\ref{fig:micro}. After 30 hours of soft etching we observe (Fig. \ref{fig:newevent}a) at high magnification, an unusual structure which is not really a track. It looks like a scratch on the detector top surface. We decided to further etch strongly the 7410-L6 layer in short time steps (5 hours) and to follow the evolution of the
``tracks'' after each etching step, systematically taking photographs. After 5 h of additional strong etching the ``tracks'' start to  
look like those in layer L1 of module 7408, as can be seen in
Fig. \ref{fig:newevent}b. After 4 and 10 more hours of strong etching
(Figs. \ref{fig:newevent}c,d) the ``tracks''  
are even more similar to those in the L1 sheet of module 7408. \par

After soft etching of the L3 and L6 layers of module 7408 we inspected all sheets in all the modules located close to 7408. We found at least one small translucent spot on one un-etched foil, suggestive of a small piece of transparent material on the surface. This background disappeared after the first soft etching. Two other surface defects were cleared in the same manner. 
In the case of three other single surface impurities, observed before 
etching in modules 7325 (L3 top), 7332 (L3 top) and 7339 (L3 top), we 
noticed that after soft and strong etching these defects yielded 
``tracks'' which resemble those seen in Fig. \ref{fig:micro}.

\section{Discussion and conclusions}

In summary, we found two strange candidate events during the analysis of the CR39 NTDs from the SLIM experiment that were exposed at the high altitude laboratory at Chacaltaya in Bolivia. 

 The first strange event candidate was observed in layer L1 of the CR39 top face in module 7408. In this case there were many etch-pits comprising the event candidate, all of abnormal shape. We made a detailed analysis of all the other layers of module 7408 and no other ``tracks'' were observed in any other layer. 

We also made a detailed analysis of all the modules which surrounded  module 7408.

 A second strange candidate event was found in the CR39 layer L6 in module 7410. Again, no more ``tracks'' were found in other layers of the module. A study of the evolution of the etch-pits with increased etching time observed in the analysis of the second candidate event casts doubts on both candidate events and lends support to the interpretation that these candidates were fakes arising from detector defects. 

This conclusion is also supported by the analysis of three other isolated single etch-pits. 

 The defects that gave rise to these strange events originated during the manufacture of the CR39 polymer. The monomer solution is poured within two casting glass plates separated by a gasket. At the end of the curing cycle, when the casting moulds are open, needle-like pieces of not perfectly cured material of microscopic size around the gasket could have been accidentally deposited on the glass moulds and become a superficial inclusion in the polymer. 
 
 The produced CR39 was laser cut into 24cm $\times$ 24cm pieces. The 
impurities may have been embedded in a number of different CR39 detectors. 

 Since the 1980's we have analyzed more than 1600 m$^2$ of CR39 using different etching conditions. During this time we did not observe any similar strange event. Thus, if these strange candidates are fakes that arise from defects in the plastic, then
we have been hit by a very rare manufacturing defect involving less than 1 m$^2$ of CR39 (at a level of less than 0.03\%). 

\section*{Acknowledgments} 

We acknowledge many colleagues and the
engineers of the CR39 manufacturers for their cooperation and technical
advices. We gratefully acknowledge the contribution of our technical
staff in Bologna and in Chacaltaya: in particular Drs. A. Casoni, M. Cecchini and R. Giacomelli. We 
thank INFN and ICTP for
providing fellowships and grants to non-Italian citizens.

\end{document}